\documentstyle[12pt,procsla]{article}
\begin{document}

\def \sp {\sqrt {16 \pi G}}
\def \ovps {\overline \psi}
\def \TI {\tilde I}
\def \DI {I^{\dagger}}
\def \TS {\tilde \sigma}
\def \DS {\sigma ^{\dagger}}
\def \Ovps {\overline \Psi}
\def \noi {\noindent}
\def \cl {\centerline}
\def \ms{\medskip}
\def \sm {\smallskip}
\def \hg {\hat g_{\hat \mu \hat \nu}}
\def \h {g_{\hat \mu \hat \nu}}
\def \g {g_{\mu \nu}}
\def \h1 {g_{ij}}
\def \f {F_{\mu \nu}}
\def \fa {F^{\mu \nu}}
\def \pa {\partial}

\centerline{\normalsize\bf EFFECTIVE WEINBERG--SALAM MODEL}

\baselineskip=22pt
\centerline{\normalsize\bf FROM HIGHER DIMENSIONS}
%\baselineskip=16pt
%\centerline{\normalsize\bf MANUSCRIPT BY COMPUTER}
%\centerline{\footnotesize\sf (For subsequent 20\% photoreduction
%to 17.8 $\times$ 11.9 cm text area)\footnote{The \LaTeX\ source
%file for this document may be used as a template for your
%article, and can be requested by e-mailing {\sf
%worldscp@singnet.com.sg}.}}

%\vfill
%\vspace*{0.6cm}
\centerline{\footnotesize Alfredo Mac\'{\i}as, Abel Camacho and Eckehard W. 
Mielke}
\baselineskip=13pt
\centerline{\footnotesize\it Universidad Aut\'onoma Metropolitana-Iztapalapa.} 
\baselineskip=12pt
\centerline{\footnotesize\it PO. Box 55-534, M\'exico D. F.09340, MEXICO.}
\centerline{\footnotesize E-mail: amac@xanum.uam.mx} 
\vspace*{0.3cm}
\centerline{\footnotesize and}
\vspace*{0.3cm}
\centerline{\footnotesize Tonatiuh Matos}
\baselineskip=13pt
\centerline{\footnotesize\it Department of Physics, CINVESTAV-IPN.}
\baselineskip=12pt
\centerline{\footnotesize\it PO. Box  14-740, M\'exico D. F. 07300, MEXICO.}
%\vfill
\vspace*{0.9cm}
\abstracts{
We consider an 8--dimensional gravitational theory, which possesses a principle
fiber bundle structure, with Lorentz--scalar fields coupled to the metric. One
of them plays the role of a Higgs field and the other one that of a dilaton 
field. 
The effective cosmological constant is interpreted as a Higgs potential. 
The Yukawa couplings are introduce by hand. The extra dimensions constitute a
$SU(2)_{L} \times U(1)_{Y} \times SU(2)_{R}$ group manifold. Dirac fields are 
coupled to the potentials derived from the metric. As result, we obtain an 
effective four--dimensional 
theory which contains all couplings of a Weinberg--Salam--Glashow theory in a 
curved space-time. The masses of the gauge bosons and of the first two 
fermion families are given by the theory.} 
 
%\vspace*{0.6cm}
\normalsize\baselineskip=15pt
\setcounter{footnote}{0}
\renewcommand{\thefootnote}{\alph{footnote}}
\section{Introduction}
Interest in higher--dimensional theories has never disappeared, it has waned 
and waxed but never has gone to zero. Recently, we\cite{1,2,3} analyzed the 
problem of the explanation for the bare mass of some 
elementary particles in the context of higher--dimensional models. 
In a further work\cite{3}, the fermionic sector of higher--dimensional 
theories was 
studied, this fermionic sector consisting of the first fundamental families. 
The space--time was endowed with the internal symmetry that corresponds to that
of the $U(1) \times SU(2)$ group with no scalar fields at all. 
Unfortunately\cite{11}, the neutrino turns out to be massive whereas the gauge
fields associated to the weak interaction are massless and the ratio between 
the leptonic and 
hadronic masses is one third, a result clearly denied by the experiment. 
It must be mentioned that the radius of the 
$S^1$ circle, is the only parameter of the extra dimensions involved in the 
masses of the fermionic sector, cf. Ref.\cite{13}, however. 

To pursue of this 
line of thought, we introduce scalar fields that could play the role
of a Higgs field and thereby could yield the terms that are necessary to solve
some of the aforementioned problems. 

To consider non--Abelian symmetries\cite{4,5} we could just take the 
structure of principal fiber bundle for the whole space--time. If we want to
introduce the electroweak interactions the structure of the principal fiber 
bundle is $(G,P) \stackrel{\pi} {\rightarrow}M^{4}$, where $M^{4}$ is the 
four--dimensional Riemannian spacetime manifold and the fiber is assumed to be
the group 
manifold of a compact non--Abelian group $G$. For the particular case that
comprises the electroweak interactions, it must be at least 
four--dimensional. When
this is combined with the four--dimensional spacetime, we are led to a 
gravity theory which is at least eight-dimensional. The line element takes the
form\cite{3,6,7}:
\begin{eqnarray}
ds^2 &=& g_{\mu \nu} (x)dx^\mu dx^\nu  - \varphi(x)^2 \left[ dx^5 + \kappa
L_1^{-1} B_\mu(x) dx^\mu \right]^2 \nonumber \\
&-& \Phi(x)^{\dagger}\Phi(x)\gamma_{ij} (y) \left[ dy^i + \kappa
L^{-1} \kappa_\alpha^i (y) A_\mu^\alpha (x) dx^\mu \right]^2
\label{1.1}\, ,
\end{eqnarray}
\noi where $\gamma_{ij}$ is the metric tensor on the group manifold of
$G$ and the functions $\kappa_\alpha^i (y)$ are the components of the
Killing vectors on $G$. The fields $A_\mu^\alpha (x)$ are gauge potentials of 
an arbitrary non--Abelian gauge group as well as components of the 
gravitational field in $4+n$ dimensions.

In the effective 4--dimensional theory, the gauge transformations are a remnant
of the original coordinate invariance group in $4+n$ dimensions, which has 
been spontaneously broken down through dimensional reduction to the symmetries
of the four--dimensional coordinate transformation group and a local gauge 
group, $\varphi(x)$ and $\Phi(x)$ are Lorentzian scalar fields, more precisely 
$\varphi$ can 
be identified with a dilatonic field, while $\Phi$ has isospin structure. 
These two type of fields depend only on the space--time coordinates. 

In the present paper we construct an eight--dimensional space--time, where
the new coordinates $y^i$ have to be interpreted as a parametrization
of the manifold of the non--Abelian group $SU(2)_{L} \times U(1)_{Y}$. 
The group manifold of $SU(2)_{L}$ is the sphere $S^3$ and that of $U(1)_{Y}$ 
is the circle $S^1$, therefore the space $S^1 \times S^3$ has the desired 
$U(1)_{Y} \times SU(2)_{L} \times SU_{R}(2)$ symmetry\cite{8} and it is the 
natural manifold for this group. In this framework we investigate an 
eight--dimensional gravity theory with two fields coupled to the metric  
(\ref{1.1}), which possesses a principle fiber bundle structure, where 
$\varphi(x)$ is 
a singlet with respect to $U(1)\times SU(2)_{L}$, namely, it is a dilaton 
field with its usual linear vacuum behavior, in which it tends to a constant 
value, but $\Phi(x)$ is now endowed with an isospin structure. Dirac fields 
are coupled to the metric one, this field contains the first two fermionic 
families, see below (\ref{3.20}.
A potential term related to the field $\Phi(x)$ that contains a mass and 
quartic self--interaction terms is introduced, it behaves as the effective 
four--dimensional cosmological constant\cite{9,10}. By hand, we introduce 
Yukawa terms, which 
consist of two contributions, namely, the bare one, which compensates the 
bare contribution of the fifth dimension that leads to non--physical results
and the usual Yukawa coupling which generates through GIM mechanism the  
fermionic masses. This means that the group structure of the right--hand part 
is $U(1)$. Through the spontaneous symmetry breaking of the 
$U(1)\times SU(2)_{L}$ symmetry of our Lagrangian and employing Weinberg 
decomposition we achieve mass terms for the gauge fields related to the weak 
interaction as well as those for the electron, muon, s, c, u, and d quarks. The
gauge field related to the electromagnetic interaction remains massless, the 
$Z$, $W^+$ and $W^-$ bosons acquire mass and their masses are in accordance 
with 
the usual relations in the four-dimensional Weinberg--Salam--Glashow theory, 
namely, the ratio between the mass of the $Z$ boson and that of the $W^+$ or 
$W^-$ is $\cos^{-2} \theta_W$, where $\theta_W$ stands for the Weinberg angle. 
Through the process of symmetry breaking the fermionic sector acquires mass, 
but there is a mass term related to both of our neutrinos that does not come 
from symmetry breaking. Again\cite{2,3} it is the influence of the fifth 
dimension, which produce the bare masses of the electron, muon, u, s, c and
d quarks.
We may conclude that for the electron, muon, $u$, $s$, $c$, and $d$ quarks, 
the mass term contains two contributions, one coming from symmetry breaking 
and the other one emerging from the presence of the radius of the fifth 
dimension, 
as a consequence of the dimensional reduction and must be cancelled by the 
Yukawa couplings, as well as for the neutrinos. 

As a consequence of dimensional reduction, in the 
four-dimensional effective theory Pauli terms arise in the usual\cite{2,3} 
manner, they may 
be understood 
as an anomalous weak momentum and an anomalous electromagnetic momentum,  
in this theory the neutrino has no electric charge, but it 
generates an electromagnetic field and this fact may be seen in the 
polarization currents that emerge in the Yang--Mills equations.  
This paper is easily extended to include the third fundamental family. 

This work is organized as follows: In section II we construct the scalar
curvature. In section III we build the eight--dimensional Dirac--Lagrangian
density. In section IV we calculate the
Yukawa couplings and afterwards carry out the breaking of symmetry. 
Section V contains Weinberg decomposition of the field 
equations by means of the mixing angle and the ensuing results are then 
discussed.

%*************************************************************
\section{Scalar curvature}

The local principal fiber bundle line element for the product space-time 
$M^4 \times G$, is given by (\ref{1.1})
where $\mu , \nu \ldots = 0,1,2,3~; i,j, \ldots = 5,6,7,8~; \alpha,\beta,
\ldots =$ group indices.
We identify $\g$ with the metric of the four--dimensional space--time,
$\gamma_{ij} (y)$ is the Killing metric on $S^1 \times S^3,~ \kappa_\alpha^i 
(y)$ are the Killing vectors and $A^\alpha_\mu(x)$ the corresponding gauge 
fields, $\varphi(x)$ is a dilaton field, while $\Phi(x)$ is an isospin 
quadruplet.  
We adopt the eight--dimensional analogue of the Einstein--Dirac--Higgs action 
\begin{equation}
I_8 = \int d^4x \int d^4y \sqrt {\hat g} \left[ {1 \over 16 \pi GV} 
\left( \hat R + V(\Phi) + Y_u  \right) + {\cal L}_D  \right] 
\label{2.2}\, ,
\end{equation}
to be the basic action. Here $\hat R$ is the eight--dimensional scalar
curvature, 
$V(\Phi)=-{\mu^2 \over 2}\Phi^{\dagger}\Phi+{\lambda \over 4!}(\Phi^{\dagger}
\Phi)^2$
is a Higgs potential term\cite{9,10},
and ${\cal L}_D$ is a straightforward generalization in eight dimensions of 
the well known four--dimensional Dirac--Lagrangian density, which will be seen 
in the next section, whereas $Y_u=h(\bar L \Phi R+\bar R \Phi^{\dagger} L)$ 
denotes the Yukawa term and $V$ is the volume of the internal space. 

We are going to employ the horizontal lift basis (HLB)\cite{6,11}
\begin{equation}
\hat \theta^{\nu} = dx^{\nu};\qquad 
\hat \theta^i = dy^i + {\kappa \over L}K^i_{\alpha}(y) A^{\alpha}_{\nu}(x)
dx^{\nu} 
\label{2.6}\, ,
\end{equation}
as basis one--forms. The basis dual to (\ref{2.6}) is. 
\begin{equation}
\hat e_{\mu}(x,y) = \partial_{\mu} - {\kappa \over L}A^{\alpha}_{\mu}
K^i_{\alpha}\partial_i;\qquad 
\hat e_i(y) = \partial_i \label{2.8}\, .
\end{equation}

On dimensional  grounds we introduce the length scales $L^{-1}$ and $L_1^{-1}$ 
of $S^3$ and $S^1$ respectively.
The metric in this basis is simply
\begin{equation}
g_{\hat \mu \hat \nu} = \pmatrix {g_{\mu \nu} (x) &0 \cr 
0 & -\h1 (y) \cr} 
\label{2.9}\, .
\end{equation}
The eight-dimensional curvature scalar is given by
\begin{equation}
\hat R = R + {R_{s^3}\over \Phi^{\dagger} \Phi}  + 2(\partial_{\nu}
\ln \varphi)
(\partial^{\nu} \ln \varphi) + {3\over \Phi^{\dagger} \Phi}(D_{\nu}
\Phi^{\dagger})(D^{\nu}\Phi)
- {\kappa^2 \over 4} (\varphi^2 F^{\mu \nu} F_{\mu \nu} + \Phi^{\dagger}\Phi 
F^{\alpha \mu \nu} F_{\alpha \mu \nu} ) 
\label{2.11}\, ,
\end{equation} 
where $R$ is the scalar curvature of $M_4$ and $R_{s^3}$ is the scalar
curvature of $S^1 \times S^3$, which is defined by
$R_{s^3} = - \gamma^{ij} R^k_{~ikj}$
so that  $R_{s^3} > 0$ for the sphere. Moreover, $g_{55}=\varphi^2 L^2_1$
and $\h1 =\Phi^2\gamma_{ij}$

%********************************************************
\section{Eight-dimensional Dirac-Lagrangian density}

As basis\cite{12,13} for the tangent bundle of $S^1$, we use the vector 
$\partial_{5}$ and for the tangent bundle of $S^3$ the three Killing vectors 
which can be written in terms of the Euler angles $(\theta, \rho, \psi)$
as follows:
\begin{equation}
{\bf K}_5 = L_1\partial_5;\qquad
{\bf K}_6 = L[\cos\psi \partial_{\theta} - \sin\psi(\cot\theta \partial_{\psi} 
- \csc\theta \partial_{\rho})] \label{3.2}\, ,
\end{equation}
\begin{equation}
{\bf K}_7 = L[\sin\psi \partial_{\theta} +\cos\psi (\cot\theta \partial_{\psi} 
- \csc\theta \partial_{\rho})];\qquad
{\bf K}_8 = L\partial_{\rho} 
\label{3.4}\, .
\end{equation}
(Note that $y^6 = \theta,~ y^7 = \rho,~
y^8 = \psi$). Thus the Killing metric $\gamma_{ij}=K_{i\alpha}K_{j}{}^{\alpha}$
for $S^1 \times S^3$ may be easily evaluated. 
We can now calculate the achtbein necessary to introduce spinors, in the HLB.
The achtbein like the metric is simply block diagonal
\begin{equation}
e^{\hat A}_{~\hat \mu} = \pmatrix{e^A_{~\mu} & 0 \cr
0 & e^{(k)}_{~~j} \cr} = \pmatrix {e^A_{~\mu} & 0 & 0 & 0 & 0\cr
0 & \varphi L_1 & 0 & 0 & 0\cr
0 & 0 & \tilde \Phi L & 0 & 0\cr
0 & 0 & 0 & \tilde \Phi L & -\tilde \Phi L \cos \theta \cr
0 & 0 & 0 & \tilde \Phi L \cos \theta & \tilde \Phi L \sin \theta \cr} 
\label{3.6}\, ,
\end{equation}
$(\hat A,\ldots ,~\hat \mu ,\ldots = 0,1,2, \ldots ,8)$ and 
$e^{\hat A}_{~\hat \mu}$ satisfy the usual relation
$\hat g_{\hat \mu \hat \nu} = e^{\hat A}_{~\hat \mu} e^{\hat B}_{~\hat \nu} 
\eta_{\hat A \hat B}$
with  $\eta_{\hat A \hat B} =$ diag $(+1, -1, \ldots,-1)$ and $e^A_{~\mu},
e^{(k)}_{~~j}$ are the vierbeins which satisfy the usual algebra
$g_{\mu\nu} = e^A_{\mu} e^B_{\nu}\eta_{AB}$;
$\gamma_{ij} = e^{(k)}_i e^{(l)}_j \delta_{kl}$
respectively. The tilde takes into account the change of sign in coupling
constants usually done in the Weinberg--Salam theory. The covariant derivative
for spinors is defined by
\begin{equation}
\hat \nabla_{\hat \mu} \psi_{jR,L} = ( \hat e_{\hat \mu} + \hat 
\Gamma_{\hat \mu} )  \psi_{jR,L} 
\label{3.10}\, .
\end{equation}
Here  
$\hat \Gamma_{\hat \mu} = {1 \over 2}\, e_{\hat A}^{~\hat \nu}\, 
e_{\hat B \hat \nu ; \hat \mu}\, \sigma^{\hat A \hat B}$ is the 
the Clifford--algebra--valued connection, with
$\sigma^{\hat A \hat B} = {1 \over 4} [ \Gamma^{\hat A}, \Gamma^{\hat B}]$.

The size of the eight-dimensional spinors is sixteen. Let $\gamma^A$
and $\gamma^k$ denote the Dirac matrices on $M^4$ and  $S^1 \times S^3$ 
respectively. Then we may take the Dirac matrices on
$M^4 \times S^1 \times S^3$ to be given by the following tensor products:
\begin{equation}
\Gamma^A = I \otimes \gamma^A \quad A= 0, 1, 2, 3\quad 
\Gamma^k = \gamma^k \otimes \hat \gamma^5 \quad k= 5, 6, 7, 8 
\label{3.13}\, ,
\end{equation}
where $\hat \gamma^5$ is the usual $\gamma^5$-matrix on $M_4$. 
The matrices in (\ref{3.13}) satisfy
$\left\{ \Gamma^{\hat A}, \Gamma^{\hat B}\right\} = 2 \eta^{\hat A \hat B}$.
Here $\gamma^A$ are the usual 4--dimensional Dirac matrices 
and $\gamma^k$ are given by
\begin{equation}
\gamma^5 = \pmatrix {i{\bf 1} & 0\cr 0 & -i{\bf 1}\cr} \quad, \quad 
\gamma^l = \pmatrix {0 & \sigma^l \cr -\sigma^l & 0\cr} \qquad l= 6,7,8
\label{3.15}\, .
\end{equation}
The eight-dimensional Dirac-Lagrangian density is defined by
\begin{eqnarray}
{\cal L}_D &=& \sum_{f=1}^2 {i \over 2} ( \overline \Psi^{fL} 
\Gamma^{\hat A}~e_{\hat A}^{~\hat \mu}~\hat \nabla_{\hat \mu} \Psi_{fL} + 
e_{\hat A}^{~\hat \mu}~\hat \nabla_{\hat \mu} \overline \Psi^{fL} 
\Gamma^{\hat A}  \Psi_{fL}) \nonumber \\
&+& \sum_{f=1}^2 {i \over 2} ( \overline \Psi^{fR} \Gamma^{\hat A}~ 
e_{\hat A}^{~\hat \mu}~\hat \nabla_{\hat \mu} \Psi_{fR} + 
e_{\hat A}^{~\hat \mu}~\hat \nabla_{\hat \mu}\overline \Psi^{fR} 
\Gamma^{\hat A}  \Psi_{fR}) 
\label{3.16}\, . 
\end{eqnarray}
Where $\Psi_{fL}$ is the left-hand part and $\Psi_{fR}$ is the right-hand 
part of our two fundamental families and the covariant derivatives for 
$\Psi_{fR}$ and $\Psi_{fL}$ are given by

\begin{eqnarray}
\nabla_{\nu} \Psi_{fR} &=& [\partial_{\nu} - {\kappa \over L_1} B_{\nu} 
\partial_{5} + \Gamma_{\nu} ] \Psi_{fR} \nonumber \\
\nabla_{\nu} \Psi_{fL} &=& [\partial_{\nu} - {\kappa \over L_{1}} B_{\nu} 
\partial_{5} - {\kappa\over L} A_{\nu}^{\alpha}K_{\alpha}^{i} \partial_{i} + 
\Gamma_{\nu}] \Psi_{fL} 
\label{3.18}\, .
\end{eqnarray}
We choose 
$\Psi_{fL} (x^\mu, y^i) = V^{-1/2} e^{iy^5Y_{L}/2} e^{i \rho \tau^3/2}
e^{i \theta \tau^1/2} e^{i \psi \tau^3/2} \psi_{fL}(x^\mu),~~(f=1,2)$
as the $x^i$ dependence for the left part of our Dirac spinor
in terms of the Euler angles in $S^3$. The left--hand part of our Dirac fields
is
\begin{equation}
\psi_{1a}(x^{\mu}) =\pmatrix{\nu_e \cr
            e^{-} \cr
            u \cr
            d} 
~,~~\psi_{2a}(x^{\mu}) =\pmatrix{\nu_{\mu} \cr 
                     \mu^{-} \cr 
                     c \cr 
                     s} 
\label{3.20}\, ,
\end{equation}
whereas the right--hand part is a singlet with respect to $SU(2)$, 
such that we may choose the following dependence for the right hand part:
$\Psi_{fR} (x^\mu, y^i) = V^{-1/2} e^{iy^5Y_{R}/2} \tilde 
\psi_{fR}(x^\mu), ~~ f=1,2$.
Here $\tilde \psi_{fR} = e_R, \mu_R, u_R, c_R, d_R, s_R$ are the right handed
singlets,
The left and right handed fermionic hypercharge matrices are $Y_{L}$ and 
$Y_{R}$ respectively. The $SU(2)$ generators $\tau^i$ are given by
\begin{equation}
Y_{\psi} = \pmatrix {1~&~0~&~0~&~0\cr 0~&~1~&~0~&~0\cr
0~&~0~&~-1/3~&~0\cr 0~&~0~&~0~&~-1/3\cr}; \quad
\tau^1 =~ \pmatrix {0~&~1~&~0~&~0\cr 1~&~0~&~0~&~0\cr
0~&~0~&~0~&~1\cr 0~&~0~&~1~&0\cr} 
\label{3.22}
\end{equation}
\begin{equation}
\tau^2 =~~ \pmatrix {1~&~-i~&~0~&~0\cr i~&~0~&~0~&~0\cr
0~&~0~&~0~&~-i\cr 0~&~0~&~i~&~0\cr} \quad~ ; \qquad
\tau^3 = \pmatrix {1~&~0~&~0~&~0\cr 0~&~-1~&~0~&~0\cr
0~&~0~&~1~&~0\cr 0~&~0~&~0~&-1\cr} 
\label{3.23}
\end{equation}
where $\tau^i$ satisfy the usual $SU(2)$ algebra
$\{ \tau^i~,~\tau^j \} = 2 \delta^{ij} {\bf 1} ; \quad
[ \tau^i~,~\tau^j ] = 2i\epsilon^{ijk} \tau_k$ 
with $Y_{R} = - 2$ for right handed leptons $e_{R}, \mu_{R}$, and
$Y_{R} = {4\over 3}$ for right handed quarks $u_{R}, c_{R}$ and 
$Y_{R} = - {2\over 3}$ for $d_{R}, s_{R}$. 
 
The form for the hypercharge matrix, $Y_{\psi}$ has been selected by 
convenience to obtain the correct values of the electric charge, using 
Gell--Mann--Nishima expression\cite{14}
$Q=T_3+{Y\over 2}$ being $Q$ the electric charge.

%**********************************************************
\section{Yukawa couplings, GIM mechanism and symmetry breaking.}

The effective four-dimensional potential is given by 
$V(\Phi) = {\mu^2\over 2} (\Phi^{\dagger} \Phi) + {\lambda\over 4!}
(\Phi^{\dagger} \Phi)^2$
where the isospin quadruplet is given by
$\Phi_{a}= \left(\phi_1\, ,\phi_2\, ,\phi_3\, ,\phi_4 \right)^{T}$.

For the hadronic part, we may introduce the 
well known Cabbibo mixture
$d_{c}=d \cos\theta_{c} + s \sin\theta_{c}$
$s_{c}=s \cos\theta_{c} - d \sin\theta_{c}$,
where $\theta_c$ is the Cabbibo angle.  
The ground state for the Higgs field is chosen as
$\Phi^0=\left(0\, ,am\, ,0\, ,an \right)^{T}$,
$\tilde \Phi^0=\left(am\, ,0\, ,an\, ,0 \right)^{T}$.
Without loss of generality we may assume $m > 0$ 
and $n> 0$, such that $n^2+m^2=1$, where $a^2 = -{6\mu^{2}/ \lambda}$. 
The covariant derivative of the $\Phi$ field is
$D_{\nu}\Phi = [\partial_{\nu} - {i\over2} g_1 B_{\nu} Y_I - {i\over 2} g_2 
A_{\nu}^{\alpha}\tau_{\alpha} ] \Phi$.

After symmetry breaking, the Yukawa terms 
for the leptons and hadrons are given as follows

\begin{eqnarray}
{\cal L} _{Yl} &=& G_1am(\bar e_L~e_R + \bar e_R~e_L) + G_2am(\bar \mu_R~\mu_L
+ \bar \mu_L~\mu_R) \nonumber \\
{\cal L} _{Yh} &=& G_5an(\bar d_L~d_R + \bar d_R~d_L) + 
G_6an(\bar s_L~s_R + \bar s_R~s_L) \nonumber \\ 
&+& G_7an(\bar u_L~u_R + \bar u_R~u_L) + G_8an(\bar c_L~c_R + \bar c_R~c_L) 
\label{4.16}\, .
\end{eqnarray}
In order to achieve the Weinberg--Salam--Glashow model in the low energy limit 
we will add to the Yukawa terms a bare contribution to compensate the 
direct influence of the fifth dimension on the fermionic masses, which leads
to results excluded by the experimental data\cite{3}.

%***********************************************************
\section{Weinberg decompositions}

We proceed to carry out the decomposition of the field equations,
in order to do so we employ the Weinberg decomposition\cite{22} 

\begin{equation}
Z_{\mu} = A^3_{\mu}c_W + B_{\mu}s_W; \quad
A_{\mu} = -A^3_{\mu}s_W + B_{\mu}c_W;\quad
W^{\pm}_{\mu} = A^1_{\mu} \mp i A^2_{\mu} 
\label{6.3}
\end{equation}
with the abbreviation $c_W = \cos\theta_W$ and $s_W =\sin\theta_W$ and 
$\theta_W$ is 
the Weinberg angle. $A_{\mu}$ is now the electromagnetic gauge field and 
$Z_{\mu}$ the neutral boson. 
Proceeding as usual in the Weinberg--Salam theory
\begin{equation}
Z_{\mu\nu} = Z_{\nu;\mu} - Z_{\mu;\nu}; \quad
A_{\mu\nu} = A_{\nu;\mu} - A_{\mu;\nu}; \quad
W^{\pm}_{\mu\nu} = W^{\pm}_{\nu;\mu} - W^{\pm}_{\mu;\nu} 
\label{6.7}\, ,
\end{equation}
the electron charge is for 
$(g_1={\kappa/ L_1}$ and $g_2 = {\kappa/ L})$ given by 
$\hat e = g_1 c_W = g_2 s_W$.
The decomposed field equations are the following:

Dirac equations:
%********************************************************
\begin{eqnarray}
&\,& i\gamma^\mu ( \partial_\mu 1_{4} + i \hat e A_\mu Q 
+ \Gamma_\mu 1_{4}) \psi_{fa} +
{1 \over 2} g_2 B_{f} + {1 \over 2} \gamma^\mu Z_\mu S_{f} \nonumber \\
&+& {i \over 2} \varphi e_A^{\mu} e_B^{~\nu} \sp ( s_W Z_{\mu \nu}
+ c_W A_{\mu \nu}) \sigma^{AB} \psi_{fa} + aM_{f} \psi_{fR} = 0 
\label{dirac}\,
\end{eqnarray}

where
\begin{equation}
Q=\pmatrix{0 \cr
            -1 \cr
            2/3 \cr
            -1/3}  
~,~~B_{1}=\pmatrix{\gamma^\mu W_\mu^+ e_L \cr 
                     \gamma^\mu W_\mu^- \nu_e \cr 
                     \gamma^\mu W_\mu^+ d_L \cr 
                     \gamma^\mu W_\mu^- u_L} 
~,~~B_{2}=\pmatrix{\gamma^\mu W_\mu^+ \mu_L \cr 
                     \gamma^\mu W_\mu^- \nu_\mu \cr 
                     \gamma^\mu W_\mu^+ s_L \cr 
                     \gamma^\mu W_\mu^- c_L} 
\label{mat}\, .
\end{equation}
\bigskip

\begin{equation}
S_{1}=\pmatrix{( g_2 c_W + g_1 s_W ) \nu_{e} \cr
            -( g_2 c_W - g_1 s_W ) e^{-}_{L} \cr
            ( g_2 c_W - {1 \over 3} g_1 s_W ) u_{L} \cr
            - ( g_2 c_W + {1 \over 3} g_1 s_W ) d_L }  
~,~~S_{2}=\pmatrix{( g_2 c_W + g_1 s_W ) \nu_{\mu} \cr
            -( g_2 c_W - g_1 s_W ) \mu_{L} \cr
            ( g_2 c_W - {1 \over 3} g_1 s_W ) c_{L} \cr
            - ( g_2 c_W + {1 \over 3} g_1 s_W ) s_L }  
\label{mat1}\, .
\end{equation}
\bigskip

\begin{equation}
M_{1}=\pmatrix{0 \cr 
                     m G_1 e_R  \cr 
                     n G_2 u_R  \cr 
                     n G_3 d_R } 
~,~~M_{2}=\pmatrix{0 \cr 
                     m G_4 \mu_R  \cr 
                     n G_5 c_R  \cr 
                     n G_6 s_R } 
\label{mat2}\, .
\end{equation}
and
$\psi_{1R}(x^{\mu}) =\left(0\, ,e_{R}\, ,u_{R}\, ,d_{R}\right)^{T}$, 
$\psi_{2R}(x^{\mu}) =\left(0\, ,\mu_{R}\, ,c_{R}\, ,s_{R}\right)^{T}$. 

For the right hand singlets: 
\begin{eqnarray}
&\,& i\gamma^\mu ( \partial_\mu 1_{4} + i \hat e A_\mu Q 
+ \Gamma_\mu 1_{4}) \psi_{fR} 
+ g_{1}s_{W}\gamma^\mu Z_\mu R_{f} \nonumber \\
&+& {i \over 2} \Phi_{1} e_A^{\mu} e_B^{~\nu} \sp ( s_W Z_{\mu \nu}
+ c_W A_{\mu \nu}) \sigma^{AB} \psi_{fR} + aM_{A} \psi_{fa} = 0 
\label{dirac1}\,
\end{eqnarray}
where

$R_{1}(x^{\mu}) =\left(e_{R}\, ,-(1/6)u_{R}\, ,-(1/6)d_{R}\right)^{T}$, 
$R_{2}(x^{\mu}) =\left(\mu_{R}\, ,-(1/6)c_{R}\, ,-(1/6)s_{R}\right)^{T}$. 

%*********************************************************

The Yang-Mills equations in the lower energy limit 
(where $\varphi=\varphi_{0}$, 
see below) are the following:

for photon:

\begin{eqnarray}
&\,&A^{\mu\lambda}_{;\mu} + 2\hat e \lbrace i(W^{[+ (\mu})_{; \mu}
W^{-]\lambda)} + {i\over 2} \lbrack W^{+ \mu \lambda}W^{-}_{\mu} +
W^{- \mu \lambda}W^{+}_{\mu} \rbrack \nonumber \\
&-& {\hat e\over 2}[({c_{W}\over s_{W}}Z^{\mu} - A^{\mu})\lbrack (W^{+\lambda}
+ W^{-\lambda})(W^{+}_{\mu} + W^{-}_{\mu}) + i(W^{+\lambda} - W^{-\lambda})
(W^{+}_{\mu} - W^{-}_{\mu})\rbrack \nonumber \\ 
&-& ({c_{W}\over s_{W}}Z^{\lambda} 
- A^{\lambda}) \lbrack (W^{+\mu} + W^{-\mu})^2 - i(W^{+\mu} 
- W^{-\mu})^2\rbrack]\rbrace = {\hat e\over \varphi^2}~\lbrack 
\bar e_{L} \gamma^{\lambda} e_{L} - {2\over 3} \bar u_{L} \gamma^{\lambda} 
u_{L} \nonumber \\
&+& {1\over 3} \bar d_{L} \gamma^{\lambda} d_{L} + \bar \mu_{L} 
\gamma^{\lambda}\mu_{L} - {2\over 3} \bar c_{L} \gamma^{\lambda} c_{L} 
+ {1\over 3} \bar s_{L}\gamma^{\lambda} s_{L} + \bar e_{R} \gamma^{\lambda}
e_{R} - {2\over 3} \bar u_{R} \gamma^{\lambda} u_{R}
+ {1\over 3} \bar d_{R} \gamma^{\lambda} d_{R} \nonumber \\
&+& \bar \mu_{R} \gamma^{\lambda} \mu_{R} - {2\over 3} \bar c_{R} 
\gamma^{\lambda} c_{R} + {1\over 3} \bar s_{R} \gamma^{\lambda} s_{R} 
\rbrack + \sqrt {16 \pi G}{c_{W}\over \varphi}[(\bar \nu_e \sigma^{\mu\lambda} 
\nu_e + \bar e_{L} \sigma^{\mu\lambda} e_{L}  \nonumber \\ 
&-& \bar u_{L} \sigma^{\mu\lambda} u_{L} - \bar d_{L} \sigma^{\mu\lambda} d_{L}
+ \bar \nu_{\mu} \sigma^{\mu\lambda} \nu_{\mu} 
+ \bar \mu_{L} \sigma^{\mu\lambda}\mu_{L} - \bar c_{L} \sigma^{\mu\lambda} 
c_{L} - \bar s_{L} \sigma^{\mu\lambda} s_{L})]_{;\mu} 
\label{6.22}\, ,
\end{eqnarray}

for the $Z_{\mu}$--boson:

\begin{eqnarray}
&\,&Z^{\mu\lambda}_{;\mu} - 2 g_{2}c_{W} \lbrace 
i(W^{[+ (\mu})_{; \mu} W^{-]\lambda)} + {i\over 2} \lbrack 
W^{+ \mu \lambda}W^{-}_{\mu} + W^{- \mu \lambda}W^{+}_{\mu} \rbrack 
\nonumber \\ 
&-& {g_{2}c_{W}\over 2}[(Z^{\mu} - {s_{W}\over c_{W}}A^{\mu})\lbrack 
(W^{+\lambda} + W^{-\lambda})(W^{+}_{\mu} + W^{-}_{\mu}) + i(W^{+\lambda} 
- W^{-\lambda})(W^{+}_{\mu} - W^{-}_{\mu})\rbrack \nonumber \\ 
&-& (Z^{\lambda} - {s_{W}\over c_{W}}A^{\lambda}) \lbrack (W^{+\mu} 
+ W^{-\mu})^2 - i(W^{+\mu} - W^{-\mu})^2\rbrack]\rbrace
+ {3a^2\over 16\pi G}[{g_1^2\over \varphi^2} + g_2^2]Z^{\mu} \nonumber \\
&=& {1 \over 2I_{1}^2}\lbrack (g_{1}s_{W} 
+ g_{2}c_{W}) (\bar \nu_e \gamma^{\lambda}\nu_e 
+ \bar \nu_{\mu} \gamma^{\lambda} \nu_{\mu}) + (g_{1}s_{W} - g_{2}c_{W})
(\bar e_{L} \gamma^{\lambda} e_{L} + \bar \mu_{L} \gamma^{\lambda} \mu_{L})
\nonumber \\
&+& (g_{2}c_{W} - {1\over 3}g_{1}s_{W}) (\bar u_{L} \gamma^{\lambda} u_{L} 
+ \bar c_{L} \gamma^{\lambda} c_{L}) 
- (g_{2}c_{W} + {1\over 3}g_{1}s_{W}) (\bar d_{L} \gamma^{\lambda} d_{L} + 
\bar s_{L} \gamma^{\lambda} s_{L})\rbrack \nonumber \\ 
&+& g_{1}s_{W} (\bar e_{R} 
\gamma^{\lambda} e_{R} + \bar \mu_{R} \gamma^{\lambda}\mu_{R})
- {1\over 3} g_{1}s_{W} (\bar u_{R} \gamma^{\lambda} u_{R} 
+ \bar c_{R} \gamma^{\lambda} c_{R}) - {1\over 3} g_{1}s_{W} (\bar d_{R} 
\gamma^{\lambda} d_{R} \nonumber \\ 
&+& \bar s_{R} \gamma^{\lambda} s_{R})\rbrack   
+ {\sqrt{16\pi G}\over \varphi}s_{W}[(\bar \nu_e \sigma^{\mu\lambda} \nu_e 
+ \bar e_{L} \sigma^{\mu\lambda} e_{L} - \bar u_{L} \sigma^{\mu\lambda} u_{L} 
- \bar d_{L} \sigma^{\mu\lambda} d_{L} \nonumber \\
&+& \bar \nu_{\mu} \sigma^{\mu\lambda} \nu_{\mu}  
+ \bar \mu_{L} \sigma^{\mu\lambda}\mu_{L} - \bar c_{L} \sigma^{\mu\lambda} 
c_{L} - \bar s_{L} \sigma^{\mu\lambda}s)]_{;\mu}
\label{6.23}\, ,
\end{eqnarray}         

for the $W^{+}$--boson:

\begin{eqnarray}
&\,&W^{+\mu\lambda}_{;\mu} - ig_{2} \lbrace \lbrack 2 W^{-[\lambda}
(c_{W}~Z^{\mu]} - s_{W}~A^{\mu]}\rbrack_{;\mu} 
- W^{+\mu \lambda}(c_{W}Z_{\mu} - s_{W}A^{\mu \lambda}) \nonumber \\  
&+& W^{-}_{\mu}(c_{W}Z^{\mu \lambda} - s_{W}A^{\mu \lambda}) + ig_{2}\lbrack
W^{+\mu}(W_{\mu i}W_{\lambda}^{i} + W_{\mu 3}W^{\lambda 3})
+ (W_{i}^{\mu}W_{\mu}^{i} + W_{3}^{\mu}W_{\mu}^{3})W^{+}_{\lambda}\rbrack 
\rbrace \nonumber \\ 
&+& {3a^2g_2^2\over 16\pi G}W^{+\mu} = g_2[\bar e_{L} 
\gamma^{\mu}\nu_e + \bar d_{L} \gamma^{\mu}u_{L}  
+ \bar {\mu_{L}} \gamma^{\mu}\nu_{\mu} + \bar s_{L} \gamma^{\mu}c_{L}] 
\label{6.24}\, ,
\end{eqnarray}

for the $W^{-}$--boson:

\begin{eqnarray}
&\,&W^{-\mu\lambda}_{;\mu} - ig_{2} \lbrace \lbrack 2 W^{+[\lambda}
(c_{W}~Z^{\mu]} - s_{W}~A^{\mu]}\rbrack_{;\mu} 
+ W^{-\mu \lambda}(c_{W}Z_{\mu} - s_{W}A^{\mu \lambda}) \nonumber \\  
&+& W^{+}_{\mu}(c_{W}Z^{\mu \lambda} - s_{W}A^{\mu \lambda}) + ig_{2}\lbrack
W^{-\mu}(W_{\mu i}W^{\lambda~i} + W_{\mu 3}W^{\lambda 3}) 
+ (W_{i}^{\mu}W_{\mu}^{i} + W_{3}^{\mu}W_{\mu}^{3})W^{-}_{\lambda}\rbrack
\rbrace \nonumber \\ 
&+& {3a^2g_2^2\over 16\pi G}W^{-\mu} = g_2[\bar \nu_e 
\gamma^{\lambda} 
e_{L} + \bar u_{L} \gamma^{\lambda} d_{L} + \bar \nu_{\mu}\gamma^{\lambda}
\mu_{L} + \bar c_{L} \gamma^{\lambda}s_{L}]
\label{6.25}\, .
\end{eqnarray}

Eqs. (\ref{dirac}) and (\ref{dirac1}) give the mass terms for the 
electron, muon, $u$, $d$, $c$, and $s$ quarks. There are two contributions, a 
dilatonic term\cite{3} proportional to $1/\varphi L_1$, which is a 
consequence
of the dimensional reduction and is cancelled by the Yukawa bare 
contribution and another one that emerges from symmetry breaking i.e., 
$am G_j$ or $an G_j$. 

>From Eqs.(\ref{6.22})--(\ref{6.25}) we find that the photon stays 
massless, and 
that the squared mass for the $W$ and $Z$ bosons are given by 

\begin{equation}
M^2_W = {3a^2 g^2_2 \over 16\pi G}; \quad
M^2_Z = {3a^2\over 16\pi G}\left[{g^2_1\over \varphi^2} + g^2_2\right] 
\label{6.27}\, ,
\end{equation}
respectively.  
These results entitle us to evaluate $a$ and the Yukawa constants 
$\bar G_j$.

In order to match these theoretical predictions with the experimental 
results$^{19, 23}$, $M^2_W/ M^2_Z =\cos^2 \theta_W$ has to be satisfied. This
leads us to conclude that our dilatonic field, $\varphi$ in its linear vacuum, 
must have a ground state in which it is a constant $\varphi_{0} \neq 0$, and 
$g_{\mu \nu} = \eta_{\mu \nu}, F_{\mu \nu} = 0, F^{\alpha}_{\mu \nu} = 0$ and
$T_{00} = 0$ is a minimum.  This is in fact the case according to our field 
equations; the ground state is degenerate and exists for any arbitrary 
constant value of $\varphi= \varphi_{0}$.
 
>From this point of view, we can choose an appropriate value of 
$\varphi_{0}$ in order to obtain the true boson masses; one finds:
$\varphi = 1 $.

Beside the usual terms of a Weinberg-Salam-Glashow theory in a curved 
spacetime, we find that as consequence of the dimensional reduction, there are
two anomalous momenta, one related to the electromagnetic gauge field 
$A_{\nu}$ and the other one associated to the weak gauge field 
$Z_{\nu}$\cite{3}. 
In Gaussian units these momenta have the value
$(\hbar/2c)\sp~c_W  \approx 5.9 \times 10^{-32} {\rm e~cm}$,
for the anomalous electromagnetic momentum and
$(\hbar/2c)\sp~s_W \approx 3.2 \times 10^{-32} {\rm e~cm}$
for the weak anomalous momentum. The interaction of these momenta with 
their corresponding gauge fields appear in the Yang--Mills equations 
(\ref{6.22})--(\ref{6.25}) and they produce additional polarization currents.

This model is readily extended to include the third fundamental family. 
and all our previous conclusions stand. 

In conclusion, we have shown that the standard model can be obtained from an 
eight--dimensional gravity theory taking  principal fiber bundle structure with
an enlarged Yukawa coupling and interpreting the effective cosmological 
constant as Higgs potential. The correct gauge bosons masses as well as the
fermionic ones are given by the theory. It seems to be that in order to 
introduce the gravitational intraction in a unified mathematical consistent
theory, the extra dimensions are needed.

%**************************************************************
\section{Acknowledgments}

This work was partially supported by CONACYT Grants No. 3544--E9311
and No. 1861--E9212.

\end{document}